\documentstyle[12pt,aasms4]{article}

\begin{document}

\title{Weak Charge-Changing Flow in Expanding $r$-Process Environments}
\author{Gail C. McLaughlin$^1$ and George M.
Fuller$^2$}
\affil{$^1$Institute for Nuclear Theory, University of Washington,  
Box 351550,
Seattle, WA 98195}
\affil{$^2$Department of Physics, University of California, San
Diego, La Jolla, CA 92093-0319 }
\authoremail{gfuller@ucsd.edu}

\begin{abstract}
We assess the prospects for attaining steady nuclear flow equilibrium  
in
expanding $r$-process environments where beta decay and/or neutrino  
capture
determine the nuclear charge-changing rates. For very rapid  
expansions, we find
that weak steady flow equilibrium normally cannot be attained.  
However, even
when neutron capture processes freeze out in such nonequilibrium  
conditions,
abundance ratios of nuclear species in the $r$-process peaks might  
still {\it
mimic} those attained in weak steady flow. This result suggests that  
the
$r$-process yield in a regime of rapid expansion can be calculated  
reliably only
when all neutron capture, photodisintegration, and weak interaction 
processes
are fully coupled in a dynamical calculation. We discuss the  
implications of
these results for models of the $r$-process sited in rapidly  
expanding
neutrino-heated ejecta.

\end{abstract}


\section{Introduction}

In this paper we study the influence of weak charge-changing nuclear  
reactions
(beta decay and neutrino capture) on neutron capture nucleosynthesis  
in rapidly
expanding (decompressing) media. The astrophysical site of origin of
$r$-process nucleosynthesis (or {\it rapid neutron capture}  
nucleosynthesis;
Burbidge {\it et al.} 1957; Cameron 1957) is not known with certainty  
({\it
cf.} Mathews \& Cowan 1990).  Recently, however, there has been a  
resurgence of
interest in $r$-process nucleosynthesis.  This interest stems from new
observations of heavy elements in low metallicity halo stars (Sneden  
{\it et
al.} 1996) and the possibility  that the $r$-process may be sited in
neutrino-heated supernova ejecta (Meyer {\it et al.} 1992; Woosley \&  
Hoffman
1992; Takahashi, Witti, \& Janka 1994; Woosley {\it et al.} 1994).  
These
considerations could have implications for particle physics and  
cosmology
since, if we understand the abundance yields of decompressing  
neutrino-heated
material, then it is possible that we could probe the basic  
properties of
neutrinos (Fuller {\it et al.} 1992; Qian {\it et al.} 1993; Qian \&  
Fuller
1995).

Though models of $r$-process nucleosynthesis from neutrino-heated  
ejecta are
promising, they suffer from a number of flaws. For example,  it is  
not
understood how the neutron-to-seed nucleus ratio in neutrino-heated  
supernova
ejecta can become high enough to produce an abundance pattern yield  
which will
match that of the solar system (see {\it e.g.} Hoffman, Woosley, \&  
Qian 1996;
Meyer, Brown, \& Luo 1996).  Yet it may be required that at least  
{\it some}
supernovae produce a solar $r$-process distribution, as there may be  
direct
observational evidence that even the earliest $r$-process events in  
the galaxy
produced an abundance pattern consistent with that observed in the  
solar system
(see {\it e.g.} Sneden {\it et al.} 1996). An alternative $r$-process  
site that
may not suffer from the neutron-to-seed nucleus problem is the  
decompression of
\lq\lq cold\rq\rq\ neutron matter from neutron star collisions  
(Lattimer {\it
et al.} 1977; Meyer 1989).

Despite problems with these models, it may be {\it necessary} to  
consider
$r$-process environments sited in an intense neutrino flux. It has  
been argued
that the observed solar system $r$-process abundance pattern itself  
may contain
clues that point to $r$-process nucleosynthesis occuring in an  
intense neutrino
flux (McLaughlin \& Fuller 1996) or requiring a significant neutrino  
fluence
(Qian {\it et al.} 1996; Haxton et al. 1996). 
Neutrino post-processing effects were also discussed in Meyer et al.
(1992).  The neutrino-heated supernova ejecta  
environment and
the neutron star merger site would each suggest that rapid neutron  
capture
nucleosynthesis takes place in an expanding medium and in an intense  
neutrino
flux.

In this paper we concentrate on how the nuclear flows in a rapid  
neutron
capture environment can be influenced by the interplay of weak charge  
changing
reactions and rapid material outflow. Our study is meant to extend  
the
evaluations of the \lq\lq waiting point\rq\rq\ assumption (that  
neutron
captures are balanced against photodisintegrations along an isotopic 
(constant
$Z$) chain, see {\it e.g.}, Cameron, Cowan, \& Truran 1982) to the  
unique
conditions of rapid outflow and intense neutrino flux which may  
characterize
neutrino-heated supernova ejecta or decompressing neutron star  
matter.

The intense neutrino flux provides a new wrinkle for $r$-process  
calculations:
in addition to the usual beta decay processes, electron neutrino  
$\nu_e$
captures on nuclei can change nuclear charge (Nadyozhin \& Panov  
1993; Fuller
\& Meyer 1995; McLaughlin \& Fuller 1995, 1996; Qian 1996; Qian {\it  
et al.}
1996). Further complicating matters,  neutrino capture rates can be  
position
(time) dependent, unlike beta decay rates.  In the comoving frame of  
a fluid
element rapidly receding from a neutrino source, the neutrino capture  
rate on
heavy nuclei will fall sharply with time. Indeed, it has been  
suggested that
rapid outflow may allow neutrino capture to dominate over beta decay  
at the
onset of the neutron capture epoch, yet allow beta decays to dominate  
weak
flows after neutron capture ceases in the regime where neutron-rich  
nuclei
\lq\lq decay back\rq\rq\ toward the valley of beta stability  
(McLaughlin \&
Fuller 1996; Qian 1996; Qian {\it et al.} 1996).

Weak nuclear flows help determine the final abundance ratios of  
nuclear species
in the abundance peaks. It has been argued that the solar system data  
provides
evidence that these ratios are within $20 \%$ of the predictions of
calculations based on weak steady flow (Kratz {\it et al.} 1988). In  
weak
steady flow, all abundances remain constant in time: neutron
capture/photodisintegration or $(n,\gamma )-(\gamma , n)$ equilibrium  
obtains
for the species along an isotopic $Z$ chain, and weak flows couple  
the
abundances of adjacent isotopic chains.

We can identify three regimes which characterize weak flows in  
r-process
nucleosynthesis in a rapidly expanding medium. These are: (1) the  
case where
the expansion timescale (or, more precisely the time available for
neutron capture)
for the medium is long compared to the  
weak
charge-changing timescale (given by the inverse of the sum of the  
typical
neutrino capture and beta decay rates);  
(2) the case where the expansion timescale is short  
compared to
the weak charge-changing timescale and (3) the case where these  
timescales are
comparable. We will show that weak steady  
flow
equilibrium can only be guaranteed in case (1), while it is  
impossible in case (2) and doubtful in case (3).

In Section 2 we consider these cases and discuss the prospects for  
attainment
of weak steady flow. We explicitly integrate the differential  
equations for
the abundances of the nuclear species in the neutron number $N=82$  
peak under
several restrictive assumptions, including that of instantaneous  
$(n,\gamma
)-(\gamma , n)$ equilibrium. In Section 3 we employ these  
calculations to
produce a plot suggesting which regions of expansion timescale and  
neutrino
flux parameter space might be conducive to either attaining weak  
steady flow
equilibrium or abundance ratios which mimic those derived from weak  
steady
flow. In Section 4 we assess the implications of these results for  
models of
$r$-process nucleosynthesis from neutrino-heated ejecta in general  
and wind
models of the post-core-bounce supernova environment in particular.

\section{Weak Flow}
\label{sec:weak}

Here we outline a method for evaluating the effect of the weak charge  
changing
flow in
a potential $r$-Process environment.
To start, we write down the
general set of equations governing the flow of nuclei in a medium  
which
contains neutrons and
heavy nuclei and which is set in a flux of neutrinos of all six  
species. For a
nucleus of charge $Z$, mass number $A$, and neutron number $N=A-Z$,  
the rate of
change of its abundance $Y(Z,A)$ (or, alternatively $Y(Z,N)$) will  
be,
\begin{eqnarray}
\label{eq:gen}
{dY(Z,A)\over dt} =
&& \sum_{n=0}^{\infty}{\left( Y(Z-1,A+n)  
{\left\{{\lambda}^{Z-1,A+n}_{\beta n}
+{\lambda}^{Z-1,A+n}_{\nu n}\right\}}\right) } - Y(Z,A)
\sum_{n=0}^{A-Z}{\left\{{\lambda}^{Z,A}_{\beta n}  
+{\lambda}^{Z,A}_{\nu
n}\right\} } \nonumber \\
&&+ \sum_{n=0}^{\infty}{\left( Y(Z,A+n){\lambda}^{NC}_{\nu
n}(Z,A+n)\right)}-Y(Z,A)\sum_{n=0}^{A-Z}{{\lambda}^{NC}_{\nu n}(Z,A)}  
\nonumber
\\
&& +\left\{ Y(Z,A-1)\lambda_{n \gamma}(Z,A-1) - Y(Z,A)\lambda_{\gamma
n}(Z,A)\right\} \nonumber \\
&& +
\left\{ Y(Z,A+1)\lambda_{\gamma n}(Z,A+1) - Y(Z,A)\lambda_{n
\gamma}(Z,A)\right\}.
\end{eqnarray}
In this equation the rates for charged current electron neutrino  
capture and
beta decay processes followed by the emission of $n$ neutrons  
proceeding on,
for example, nucleus $(Z,A)$ are ${\lambda}^{Z,A}_{\nu n}$ and
${\lambda}^{Z,A}_{\beta n}$, respectively. The rate of neutral  
current neutrino
scattering on, for example, nucleus $(Z,A)$ accompanied by the  
spallation
(emission) of $n$ neutrons is ${\lambda}^{NC}_{\nu n}(Z,A)$. The  
neutron
capture rate (photodisintegration rate)
on, for example,  nucleus $(Z,A)$ is labelled as $\lambda_{n  
\gamma}(Z,A)$
($\lambda_{\gamma n}(Z,A)$). Here we have ignored charged particle  
nuclear
reactions, and we take photodisintegration rates to include the  
effects of
charged particle-induced Coulomb excitation followed by neutron  
emission. The
first line of this equation represents the effects of charged current  
weak
processes, while the second line gives the effects of neutral current  
neutrino
scattering-induced neutron spallation. The third and fourth lines of  
this
equation show the effects of $(n,\gamma )$ and $(\gamma , n)$  
reactions. In
general all of the above terms need to be included in the calculation  
of
$dY(Z,A) / dt$.
However, in some conditions certain simplifying assumptions can be  
made.

For example, at sufficiently high temperature and density
($T_9 \gtrsim 1$, $n_n \gtrsim 10^{20} {\rm cm}^{-3}$, where $T_9$ is  
the
temperature in units of $10^9 \, {\rm K}$ and $n_n$ is the neutron  
number
density cf. Meyer et
al. 1992),
$(n,\gamma)-(\gamma,n)$-equilibrium obtains.  In this equilibrium  
situation,
the rate of
photodissociation reactions is balanced by
 the rate of neutron captures for a given set of nuclei.  This
often occurs in conditions when the weak charge changing rates are  
much slower
than the
neutron capture rates, so
that the waiting point approximation is valid (Cameron, Cowan and  
Truran 1982).
If $(n,\gamma)-(\gamma ,n)$ equilibrium is assumed to be established  
on a
timescale rapid compared to that of any of the weak processes, then  
we can
ignore neutral current neutrino scattering-induced neutron spallation  
and,
further, rewrite the above equation in terms of {\it inclusive}, or  
total weak
charged current rates, {\it e.g.} ${\lambda}_{\beta}(Z,N) \equiv
\sum_{n=0}^{A-Z}{{\lambda}^{Z,A}_{\beta n}} $, and  
${\lambda}_{\nu}(Z,N) \equiv
\sum_{n=0}^{A-Z}{{\lambda}^{Z,A}_{\nu n}}$. Of course, under the  
general
assumption of $(n,\gamma),(\gamma ,n)$ equilibrium,
neutron capture reaction rates will be equal and opposite to  
corresponding
photodisintegration rates, so that we can drop these terms in  
equation
(\ref{eq:gen}) to obtain,
\begin{eqnarray}
 \label{eq:spec} {dY(Z,N) \over dt} =
&& - [ \lambda_{\beta}(Z,N) + \lambda_{\nu}(Z,N)]Y(Z,N)  \nonumber \\
&& +[\lambda_{\beta}(Z-1,N+1) + \lambda_{\nu}(Z-1,N+1)]Y(Z-1,N+1).
\end{eqnarray}
The reactions governing the abundances of a set of nuclei  can be
described as a series of coupled differential equations, of the form  
of
equation
(\ref{eq:gen}).  In regions where the waiting point approximation is  
valid,
this set may be reduced to equations of the form of (\ref{eq:spec}).
The nuclei which comprise the peaks in the $r$-Process abundance  
distribution
{\it may} be an example of such a set.

Sometimes it is also possible to make the steady flow approximation.
Steady beta flow has been invoked to explain the measured solar  
system
$r$-Process abundance peaks (cf. Kratz et al. 1988).  The definition  
of steady
flow
equilibrium is
as follows: if the input to a system of nuclei, such as the nuclei  
which
comprise
the $r$-Process peaks, at the
nucleus of lowest Z, is equal to the output at the nucleus of highest  
Z, then
steady flow equilibrium obtains.  As a consequence of
steady beta flow equilibrium in an environment with no neutrinos, the  
ratio of
abundances of progenitor $r$-process nuclei in the
peaks will be the inverse ratio of the beta decay rates.
More generally with neutrinos, the ratio of these abundances will  
reflect
the inverse ratio of the neutrino capture plus beta decay rates when  
steady
{\it weak} flow equilibrium obtains.

We wish to evaluate the prospects for attaining weak steady flow  
equilibrium in
an expanding
outflow such as that of the post-core-bounce supernova.  To this end,  
we write
the formal solution
of equation (\ref{eq:spec}) for a particular $Y(Z,N)$:
\begin{eqnarray}
\label{eq:soln}
Y(Z,N,t_f) =  && \left[ Y(Z,N,t_i) - {\lambda_{\beta +
\nu}(Z-1,N+1,t_i)Y(Z-1,N+1,t_i) \over
\lambda_{\beta + \nu}(Z,N,t_i)} \right] f(t_i,t_f)
 \nonumber \\
&& + \left[ {\lambda_{\beta + \nu}(Z-1,N+1,t_f)Y(Z-1,N+1,t_f) \over
\lambda_{\beta + \nu}(Z,N,t_f)} \right] \nonumber \\
&& - \int^{t_f}_{t_i} {d \over dt}\left[{\lambda_{\beta +
\nu}(Z-1,N+1,t)Y(Z-1,N+1,t) \over
\lambda_{\beta + \nu}(Z,N,t)}\right]  f(t_f,t),
\end{eqnarray}
$$ f(a,b) = \exp{\left[- \int^{b}_{a} \lambda_{\beta+\nu}(Z,N,t) dt  
\right]} $$
Here, the time dependence of the rates and the abundances has been  
explicitly
included and the
solution is evaluated at the final time, $t_f$.  The first line shows  
the
dependence of
the solution on the initial conditions (time $t_i$).  For compactness  
of notation we have denoted
the sum $\lambda_\beta(Z,N) + \lambda_\nu(Z,N,t)$ as $\lambda_{\beta  
+ \nu}(Z,N,t)$.
The beta decay rates are independent of time,  while the time  
dependence of
the neutrino capture rates is linked to the outflow rate of the  
material.
In equation (3), the
initial time corresponds to the onset of $(n,\gamma)-(\gamma,n)$  
equilibrium
and the final time
corresponds to freeze out from $(n,\gamma)-(\gamma,n)$ equilibrium.   
It is apparent from equation
(\ref{eq:soln}) that the comparison between the total time spent in
$(n,\gamma)-(\gamma,n)$ equilibrium and the lifetime of a nucleus  
against beta
decay and neutrino capture is an important factor in determining the
abundances.  We divide the
solutions in three categories based on this comparison.

The first category is composed of the solutions for which
$\lambda_\beta(Z,N, t_f) + \lambda_\nu(Z,N, t_f) \gg 1 / (t_f -  
t_i)$.
Here the term in the first line in the above equation is very small,  
due to the large negative factor in the
exponential, so that the effect of the initial conditions has been  
erased.
In addition, the exponential in the last term is very small, except  
for the
last time interval
$ \delta t \sim 1/[\lambda_\beta(Z,N,t) + \lambda_\nu(Z,N,t)] $.
As long as the rates are large, the
range of integration where the integrand is nonnegligible is very  
small,
rendering the last term approximately equal to zero.
Therefore the ratio of abundances of $Y(Z,N)$ to $Y(Z-1,N+1)$ is  
approximately
the inverse ratio of the rates and steady weak flow is obtained.
This is the regime of applicability of the analysis of
constraints on or contributions from the neutrino flux as employed by 
Fuller and Meyer (1995) and McLaughlin and Fuller (1996).

The second category is the situation where
$\lambda_\beta(Z,N, t_f) + \lambda_\nu(Z,N, t_f) \ll 1 / (t_f -  
t_i)$.
In this case, the solution is governed mostly by the initial  
conditions, since
there is very little
time to move nuclei around with weak charge changing reactions. In  
this case,
the $r$-Process can
not take place, since the nuclei will not be able to move up the  
proton number
ladder fast enough
to produce the peaks seen in the $r$-Process abundance measurements.

The third category is the intermediate regime, where the solution
is not entirely governed by the initial
conditions and yet steady flow has not been attained.  There is a  
fairly wide
range of conditions
around $\lambda_\beta(Z,N, t_f) + \lambda_\nu(Z,N, t_f) \sim 1 / (t_f  
- t_i)$,
which fall into this
category.  In this case all the terms in equation (\ref{eq:soln})   
must be
included in order to
predict the abundance of element $Y(Z,N)$ at the time of freeze out  
from
$(n,\gamma)-(\gamma,n)$ equilibrium.  As discussed later in this  
section,
despite the fact that steady weak flow has not been attained, it
is possible in some cases to mimic the ratios expected in steady beta  
flow
conditions.

In order to give an example of
the third intermediate case, we present the solution for the $N=82$  
peak
for a particular set of conditions.  We integrate forward
four coupled differential equations of the type of
equation (\ref{eq:spec}), corresponding to each of the nuclei,  
$^{127}{\rm
Rh}$, $^{128}{\rm Pd}$,
$^{129}{\rm Ag}$, and $^{130}{\rm Cd}$.  The beta decay rates and  
neutrino
capture rates are taken
from Table 1 of McLaughlin and Fuller (1996).  The neutrino capture  
rates scale
as $\propto 1/r^2$,
where r is the distance from the neutrino sphere.
Therefore, rapidly outflowing environments will produce rapidly
changing neutrino capture rates.  We take the the distance coordinate  
r to vary
as $r \propto
\exp(t/\tau_e)$, where $\tau_e$ is the expansion timescale.  We make  
the
simplifying assumption
that there is a constant input at the bottom of the ladder (at   
$^{127}{\rm
Rh}$).  Note that
in the
true $r$-Process environment, it may be necessary to describe the  
input as a
more complicated
function of time.  In fact, the initial distribution of seed nuclei
in the N, Z - plane at the onset of neutron capture, and the time 
dependence of the thermodynamic conditions can influence nuclear flow.
Thus our calculations are quite  
schematic and
are designed only to illustrate salient physical effects. We take the
abundances of all
the nuclei to be zero at the beginning of our calculation.  In  
addition, we
begin the integration
at a time when the material is
close to the neutrino sphere, so that the neutrino capture rates are  
roughly
ten times the beta
decay rates at the onset of $(n,\gamma)-(\gamma,n)$ equilibrium

The solution is plotted in Figure (\ref{fig:fig1}) as the ratio of  
$^{130}{\rm
Cd}$
to $^{127}{\rm Rh}$ against time.  This lower curve
at first rises rapidly, since the fast neutrino capture rates are  
causing the
nuclei to climb the
ladder to  $^{130}{\rm Cd}$.  However, after a few tenths of a second  
the curve
turns over,
due to the rapidly falling neutrino capture rates.  The abundance  
ratios
can not keep pace with the
changing neutrino capture rates, and nuclei begin to pile up at the  
bottom of
the ladder.  After a
little more than a second, the curve turns upward again.  This  
signals the
point at which the beta
decay rates become comparable to the neutrino capture rates.  The  
solution then
asymptotes to the
steady beta flow value.  Note that in an environment where the  
neutrino capture
rates are small, the
solution would begin at zero and slowly asymptote to the steady beta  
flow value
(i.e. there would be no spike and subsequent turnover at the early  
time).

For comparison, the inverse ratio of the weak rates is also plotted  
against
time (upper curve) in this figure.
When the calculation begins, neutrino captures dominate over beta  
decays and
the ratio of the rates is high.  This is due to the small variation  
of neutrino
capture rates
between nuclei of similar neutron and proton number (Fuller \& Meyer  
1995,
McLaughlin \& Fuller
1995).  At late times, beta decays dominate over neutrino captures.   
Since beta
decay rates show
larger variation between nuclei, the ratio is significantly different  
from
unity at late times.

It is clear from the figure that for this choice of conditions,
steady flow is not attained until a
few seconds has elapsed.
In order to determine the abundance ratios it is necessary to locate  
the time
of
freeze out from of $(n,\gamma)-(\gamma,n)$ equilibrium on Figure
\ref{fig:fig1}.
For example, in the wind model which is discussed by  Qian et al.  
(1996),
freeze out is at
$1.1\tau_e$.  For our example parameters, that occurs at 0.55  
seconds.
At this time, the value of the abundance ratios is
within 20\% of the steady beta flow values, despite the fact that  
steady beta
flow has not been
attained.  In fact, the abundance ratio is determined by the  
expansion
timescale and the overall
magnitude of the neutrino capture rates and has very little
to do with the values of the beta decay rates for these nuclei.
This example serves to demonstrate that it is possible
to mimic steady beta flow ratios when the neutrino capture rates are  
dominant
or at least
comparable to the beta decay rates.  Such a scenario is highly model  
dependent
and, due to the
large contribution from neutrino capture, will usually necessitate  
significant
neutrino post
processing effects. Our simple calculation cannot, of course, obtain  
the the
details of $r$-process nucleosynthesis. It can suggest only where it  
will be
necessary to perform fully coupled calculations which do not rely on
equilibrium assumptions.

\section{Weak Flow in Wind-like Models}

In this section we explore the range of neutrino flux conditions and
expansion timescales which
may produce steady weak flow, steady beta flow, or abundances which  
mimic
steady weak flow
in a rapidly expanding environment.
We show the link between these parameters and get an estimate of the  
degree to
which the
$r$-Process progenitor abundances
will experience neutrino post processing.  In order to investigate  
the
available
parameter space, it is necessary to adopt a model which relates the  
timescale
over which significant neutron capture occurs to the material  
expansion
timescale. In the context of the simple model employed above, the  
time during
which neutron capture occurs will be taken to be the time over which  
$(n,\gamma
)-(\gamma ,n)$ equilibrium obtains.  Ideally, an analysis
of weak flow should be conducted with a sophisticated  
multi-dimensional
hydrodynamic model for  the post-core-bounce supernova outflow or the  
neutron
star collision scenario which makes no recourse to equilibrium  
assumptions.
Although understanding of multidimensional Type II supernova models has greatly
improved recently, they do not yet give a definitive picture of the late time
outflow conditions during which $r$-Process nucleosynthesis may take place 
(Herant et al 1992, 1994, Miller et al. 1993, Burrows et al. 1995, Janka \&
M\"uller 1996, Mezzacappa et al. 1996).
In light of this, we will employ the simplistic
$(n,\gamma )-(\gamma, n)$ equilibrium calculation of section  
\ref{sec:weak} in
the context of one-dimensional steady-state wind models. This will  
facilitate
exploration
of the weak flow parameter space and elucidate important physics  
issues for $r$-Process nucleosynthesis.
Neutrino-driven wind models have in the past proved useful for studying 
other aspects of the post-core-bounce environment such as the neutron-to-seed 
ratio and neutrino-induced neutron spallation during post-processing
(Qian \& Woosley 1996; Hoffman, Woosley \& Qian 1996; Qian et al. 1996).

Figure (\ref{fig:fig2}) shows a plot, specific to the wind model,
of possible solutions to the set of
coupled differential equations which govern weak flow in the $N=82$  
peak.  The
vertical axis is
$(L_\nu / 10^{51} {\rm ergs}) (100 \, {\rm km} / r)^2$ evaluated at
freeze out from $(n,\gamma)-(\gamma,n)$ equilibrium. Here $L_\nu$ is  
the
neutrino
luminosity and $r$ is the distance from the neutrino
sphere.  The horizontal axis is the expansion timescale (i.e. the  
dynamical
time $\tau_D$ in
the wind model).
The abundance ratios for each solution were evaluated at the time of
 $(n,\gamma)-(\gamma,n)$ freeze out and compared to those that
should obtain in steady weak flow and steady beta flow.
Note that the neutrino capture rates are roughly
comparable to the beta decay rates when
$\log_{10} \left[ (L_\nu / 10^{51} {\rm ergs}) (100 {\rm km} / r)^2  
\right]
\sim 1$.

The first question we wish to answer is: for what range of parameters  
does
steady weak flow
obtain?  We choose the criterion for defining
steady weak flow to be when the output from the top of the $N=82$  
peak
chain is with 10\% of the input at the bottom of the chain, evaluated  
at the
time of
$(n,\gamma)-(\gamma,n)$ freeze out.  On the graph there are
two separate regions of parameter space where steady weak flow  
obtains.  One
area resides in the upper right hand corner.  In this region the  
neutrino
capture rates are fast and
the expansion timescale is slow, so that steady weak equilibrium is  
established
quickly.  In the shaded region in the lower right hand corner steady  
beta flow
obtains,
a limiting case of steady weak flow.  In this region, the neutrino
capture rates are small enough that they have very little impact on  
the
solution.  It is
interesting to note that a very large or very small neutrino flux  
assists in
creating
steady weak flow, while a moderate amount can prevent its  
establishment in a
\lq\lq wind-like\rq\rq\ exponential expansion.  This is due to
the inability of the abundances ratios to keep pace with the rapidly  
changing
neutrino capture
rates.
It is evident from Figure (\ref{fig:fig2})
that steady weak flow is not a good approximation for describing
the weak charge changing flow in a fast expansion wind model. Indeed,   
in some
senses the most natural wind models have dynamical times of $0.1 -  
0.2$ seconds
(Qian \& Woosley 1996; Duncan, Shapiro and Wasserman 1986) 
and for these steady flow equilibrium is clearly a  
bad
approximation.

The second question involves the mimicry of steady beta flow  
abundance ratios.
As
mentioned in
section \ref{sec:weak}, the abundance ratios may be close to their  
steady beta
flow values,
despite the fact that the system is
not in steady beta flow or even the more general steady weak flow.
In Figure (\ref{fig:fig2}) the region between the thick solid
lines is the region where all of the abundance ratios in the $N=82$  
peak at
neutron capture
freeze out in our simplistic calculations will be within 20\% of the  
values
that would obtain if the system were in
steady beta flow.

We have also placed a charge changing constraint line
on this plot for comparison.  If there are
roughly four charge changing reactions that must take place in order  
for a
nucleus to travel up
the proton number ladder in the $N=82$ peak, then this places a  
constraint on
the expansion
timescales (see for example Qian 1996).  On the figure, above the  
long dashed
line, a nucleus at the bottom of the $N=82$
peak will experience more than four charge changing reactions before
$(n,\gamma),(\gamma,n)$
freeze-out.

Since the region of steady beta flow mimicry occurs when there is a  
significant or even
dominant neutrino capture component to the total charge changing rate
at $(n,\gamma)-(\gamma,n)$ freeze-out, it behooves us to ask
how much neutrino post processing will occur in this range of  
parameters.
Several neutrons will be emitted after a
neutrino capture reaction, since such a reaction leaves the daughter  
nucleus
in a highly excited state.  If there are a significant percentage of  
nuclei
experiencing such
interactions,
then the decay back to the valley of beta stability after
$(n,\gamma)-(\gamma,n)$ freeze out
will be significantly altered from the traditional picture.  In fact,  
if all
the
nuclei experience one
or more neutrino captures, the peak will be shifted toward a region
with lower total nucleon number.
It is possible to
integrate the weak charge changing rates to determine the percentage  
of nuclei
that will
experience a neutrino capture (Qian et al. 1996).   Here for  
illustration
we only include the contribution of the charged current processes;  
neutral
spallation processes would give
an additional (possibly comparable) contribution.   In order to  
illustrate the post processing effects,
we assume that the expansion timescale remains the {\it same}  
throughout the
period of $(n,\gamma)-(\gamma,n)$ equilibrium and the neutrino post  
processing
epoch. With a
timescale of a tenth of a second, which is a value typically  
discussed for the
wind model, the region of mimicry
occurs at a sufficiently high neutrino flux that almost all nuclei  
will
experience one
post-processing charged current neutrino interaction.  Although a  
detailed
calculation is necessary to determine the
exact abundances that would be produced by this model, it is clearly  
unlikely
that the observed $r$-process distribution can be reproduced with a  
single wind expansion timescale of a tenth of a second.

In Figure (\ref{fig:fig2}), the region where nuclei will
experience less than one neutrino capture per nucleus is shown as the  
area
below the upper
short dashed line.  The region where less than 25\% of the nuclei  
will
experience a neutrino
capture is below the lower short dashed line.  There is only a very  
small
region in the single expansion timescale wind model
where postprocessing effects are small and steady beta flow effects  
might be
mimicked.  This occurs in
the small unshaded triangle on the vertical axis of the graph,
marked by the short dashed line and the heavy solid line.  This is  
probably the
best scenario for
reproducing the $r$-process abundances with a single timescale  
exponentially
expanding wind model.  Only a detailed network calculation employing  
the full
machinery of Equation 1 could resolve this conjecture. Even so, this  
scenario
employs an expansion timescale of one second. This timescale is  
considerably
larger than that usually discussed for the wind model, but could be  
obtained
conceivably given uncertainties in the neutrino emission and  
hydrodynamics in
the post-core-bounce supernova environment.

\section{Conclusions}

In this paper we have presented a guide for analyzing the weak  
charged changing
flow in a potential
$r$-Process environment.  We have identified three different regimes  
for the
weak flow,
which depend on the neutrino flux and the expansion timescale.   
Steady weak
flow
occurs when the time spent in
$(n,\gamma)-(\gamma,n)$ equilibrium is very long when compared with  
the
inverse charge changing rate.

In order to asses the impact of these weak flow considerations on the
nucleosynthesis yield in a potential $r$-Process site, in section 3  
we
applied the weak flow analysis to a model with an exponential  
outflow.
We concentrated on the $N=82$ peak, although the same analysis may  
also be
applied
to other regions such as the $N=126$ peak. (Clearly, the favored  
parameter
space of expansion timescale and neutrino flux will be somewhat  
different for
the material which at the conclusion of neutron capture makes the $A  
= 195$
material.)  We find that
the material is only in weak steady flow equilibrium
when the expansion timescale is very long, such as a few seconds.
At the much shorter times favored in the wind model, such as a tenth  
of a
second, the
$r$-process abundances can still conceivably mimic those of steady  
beta flow.
However, these probably finely tuned scenarios might
necessitate neutrino-induced neutron spallation post-processing  
during the
decay back to beta stability at a level which could be intolerable.
A full network calculation carried through the period of decay back  
to beta
stability
would be necessary, however, to answer this question definitively.

{}From our analysis of steady weak flow, we can infer the sort of  
conditions
which would be
conducive to producing the peaks in the $r$-process abundance  
distribution.
One possibility is that neutrino capture has a limited impact
during the neutron capture phase.  Then the potential environments  
are
restricted to those
where steady beta flow obtains or nearly obtains.
One option might be a
slightly altered version (altered to produce the correct  
neutron-to-seed ratio)
of the slow one dimensional post-core-bounce outflow produced in the  
Type II
supernova model
of Mayle and Wilson (as employed in the Meyer {\it et al.} 1992 and  
Woosley
{\it et al.} 1994 $r$-process calculations).
Another possibility is that the
 neutrino captures play a major role in accelerating the
weak charge changing flow during the
neutron capture phase, but the expansion
rate takes on a more complicated time dependence than the single  
timescale
exponential wind model employed here.
An example of such a scenario occurs when the expansion rate is fast
during alpha rich freeze out so that the neutron-to-seed ratio is  
acceptable
({\it e.g.} Hoffman, Woosley, \& Qian 1996),
slow during the period of $(n,\gamma)-(\gamma,n)$-equilibrium  
(neutron capture
epoch ) so
that steady weak flow can obtain, and then relatively fast during the
post-processing phase so
that the amount of neutron spallation is restricted.   Such a \lq\lq
fast-slow-fast\rq\rq\  scenario sounds excessively
convoluted in the one-dimensional post-core-bounce outflow models,  
but it
remains to be seen whether
three dimensional convective models could produce such conditions  
(envision initially
rapidly outflowing material which during the epoch of neutron capture 
is caught in a  
convective eddy
and then later re-accelerated and ejected).  
A third possibility is that the
site of $r$-Process nucleosynthesis is not the post-core-bounce  
supernova
environment. We caution, however, that decompression of cold neutron  
matter
from neutron star mergers/collisions could well experience the same  
parameter
space \lq\lq squeeze\rq\rq\ we describe here for wind models of  
neutrino-heated
supernova ejecta.

We conclude by emphasizing that only a full network calculation,  
including the
neutron capture period and the post processing
period can predict accurately the nucleosynthesis yield which results  
from a
rapidly
expanding environment.

\acknowledgments
We wish to thank Y.-Z. Qian and W. C. Haxton for useful discussions.
This work was supported by NSF Grant PHY-9503384 at UCSD, and by  
Department of
Energy Grant DE-FG06-96ER40561 at the INT.

\begin{figure}
%\plotone{fig1.ps}
\caption
[The abundance ratio of  $^{130}{\rm Cd}$ to $^{127}{\rm Rh}$ is plotted against time.]
{ \label{fig:fig1}
The abundance ratio of $^{130}{\rm Cd}$ to $^{127}{\rm Rh}$ is plotted against time.  Also shown
for comparison is the inverse ratio of the weak rates (beta decay  plus neutrino capture) for these
nuclei.  If at any time the two curves took on the same, or a similar value, then weak steady flow
would obtain.  In this plot, the distance from the neutrino sphere, $r$ depends on time, $t$ as 
$r \propto \exp(t/\tau_e).$ For illustrative purposes we have taken the expansion time, 
$\tau_e$ to be $0.5 \, {\rm s}$.  When the abundance ratio enters the region between the two
dashed lines, then the abundance ratio is within 20\% of the value it would take on in conditions 
of
steady beta flow.}
\end{figure}
\begin{figure}
%\plotone{fig2.ps}
\caption
[This plot delineates the parameter space available in the wind model.]
{\label{fig:fig2}
This plot delineates the parameter space available in the wind model.  The quantity $(L_\nu /
10^{51} {\rm erg \, s^{-1}}) (100 \, {\rm km}  / r)^2$ evaluated at 
$(n,\gamma)-(\gamma,n)$ equilibrium freeze-out is 
plotted against
dynamical time.  Steady beta (weak) flow equilibrium obtains in the shaded region in the lower 
(upper) right 
hand corner.  In the region between the dark solid lines, the abundance ratios mimic
steady beta flow values.  In the region below the long dashed line, there is not enough 
time for a nucleus to
change sufficient charge to traverse the N=82 peak.  In the region above the upper short dashed
line, there is more than one post processing neutrino capture per nucleus.  In the region above the
lower short dashed line, more than 25\% of the nuclei experience a post processing neutrino 
capture.}
\end{figure}

\begin{references}

\reference{BBFH57}
Burbidge, E. M., Burbidge, G. R., Fowler, W. A., \& Hoyle, F. 1957,  
Rev. Mod. Phys., 29, 624

\reference{BU95}
Burrows, A., Hayes, J., \& Fryxell, B. A., 1995, ApJ, 450, 830

\reference{Cam57}
Cameron, A. G. W. 1957, Stellar Evolution, Nuclear Astrophysics, and
Nucleogenesis (Chalk River Rep. CRL-41) (2d ed.; Chalk River,  
Ontario: At.
Energy Canada Ltd.)

\reference{CCT83}
Cameron, A. G. W., Cowan, J. J. , \& Truran, J. W. 1983, Ap\& SS, 91,  
235

\reference{CTT91}
Cowan, J. J., Thielemann, F.-K., \& Truran, J. W. 1991, Phys. Rep.,  
208, 267

\reference{DU86}
Duncan, R. C., Shapiro, S. L., \& Wasserman, I. 1986, ApJ, 309, 141

\reference{FM92}
Fuller, G. M., Mayle, R., Meyer, B. S., and Wilson, J. R., 
1992, ApJ, 389, 517

\reference{FM95}
Fuller, G. M., and Meyer, B. S. 1995, ApJ, 453, 792

\reference{HA96}
Haxton, W. C., Langanke, K., Qian, Y.-Z., and Vogel, P. 1996, preprint

\reference{HE94}
Herant, M. E., Benz, W., Hix W. R., Fryer, C., \& Colgate, S. A. 1994, ApJ, 435,339,

\reference{HE92}
Herant, M. E., Benz, W. \& Colgate, S. A. 1992, ApJ, 395, 642

\reference{HO96}
Hoffman, R. D., Woosley, S. E., \& Qian, Y.-Z., 1996, ApJ, submitted

\reference{JA96}
Janka, H.-Th., \& M\"uller, 1996, A\& A, 306, 167

\reference{KR88}
Kratz, K.-L., Bitouzet, J. P., Thielemann, F.-K., M\"oller, P., and Pfieffer, B. 
1993, 403, 216

\reference{KR93}
Kratz, K.-L., Thielemann, F. K., Hillebrant, W., M\"oller, P., H\"arms, V., W\"ohr, 
A., \& Truran, J. W. 1988, J. Phys. G 24, S331

\reference{LA77}
Lattimer, J. M., Mackie, F., Ravenhall, D. G. \& Schramm, D. N., 1977, ApJ, 213, 
225

\reference{MA90} 
Mathews, G. J., \& Cowan, J. J., 1990, Nature, 345, 491

\reference{MF95}
McLaughlin, G. C. \& Fuller, G. M. 1995, ApJ, 455, 202

\reference{MF96}
McLaughlin, G. C. \& Fuller, G. M. 1996, ApJ, 464, L143

\reference{ME89}
Meyer, B. S., 1989, ApJ, 343, 254

\reference{ME96}
Meyer, B. S., Brown, J. S. \& Luo, N. 1996, preprint.

\reference{ME92}
Meyer, B. S., Howard, W. M., Mathews, G. J., Woosley, S. E., 
\& Hoffman, R. D. 1992, ApJ, 399, 656

\reference{ME96}
Mezzacappa, A., Calder, A. C., Bruenn, S. W., Blondin, J. M.,  
Guidry, M. W., Strayer, M. R., \&
Umar, A. S. 1996, ApJ, submitted

\reference{MI93}
Miller, D. S., Wilson, J. R. \& Mayle, R. W. 1993, ApJ, 415,278

\reference{NaPan93}
Nadyozhin, D. K., \& Panov, I. V. 1993, in Proc. Int. Symp. on Weak  
and
Electromagnetic Interactions in Nuclei (WEIN-92), ed. Ts. D. Vylov  
(Singapore:
World Scientific), 479

\reference{QI93}
Qian, Y.-Z., Fuller, G. M., Mathews, G. J., Mayle, R. W., Wilson, J. R. \& Woosley, S. E., 1993,
Phys. Rev. Lett., 71, 1965

\reference{QI95} Qian, Y.-Z. \& Fuller, G. M. 1995, Phys. Rev. D, 52, 656.

\reference{QI96}
Qian, Y.-Z., Haxton, W. C., Langanke, K. and Vogel, P., 1996, submitted to Phys Rev. C.

\reference{QI96a}
Qian, Y.-Z., 1996, Invited Talk at the Fourth International Conference on Nuclei in the Cosmos,
University of Notre Dame, USA

\reference{QI96b}
Qian, Y. Z., 1996 \& Woosley, S. E., 1996, 471, 331

\reference{SN96}
Sneden, C., McWilliam, A., Preston, G. W., Cowan, J. J., ApJ, 1996, 467, 819

\reference{TA94}
Takahashi, K., Witti, J. \& Janka, H.-Th., 1994, A \& A 286, 857

\reference{WO92} 
Woosley, S. E. \& Hoffman, R. D. 1992, ApJ, 395, 202

\reference{WO94}
Woosley, S. E., Wilson, J. R., Mathews, G. J., Hoffman, R. D., 
\& Meyer, B. S., 1994, ApJ, 453, 229.

\end{references}
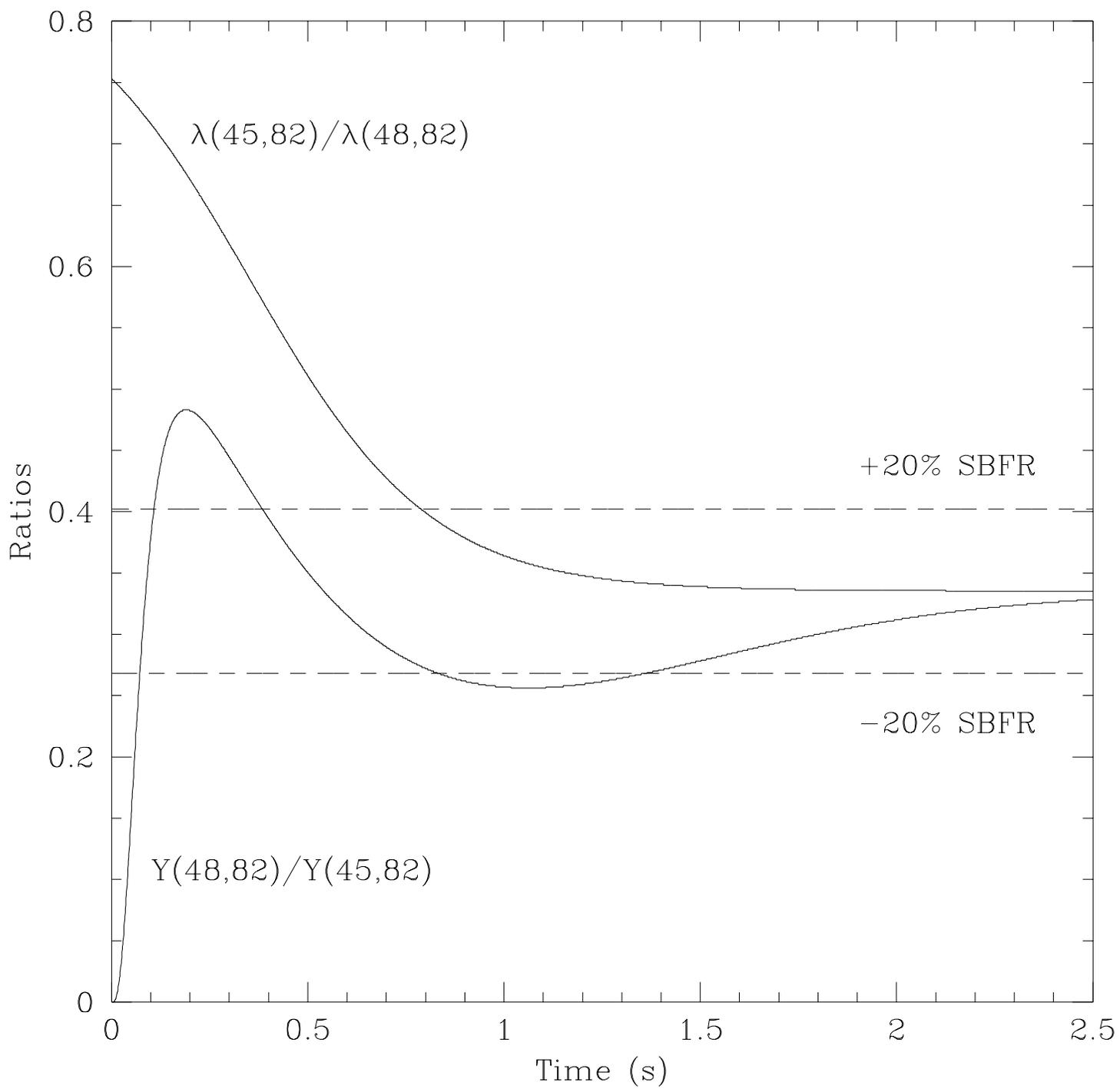
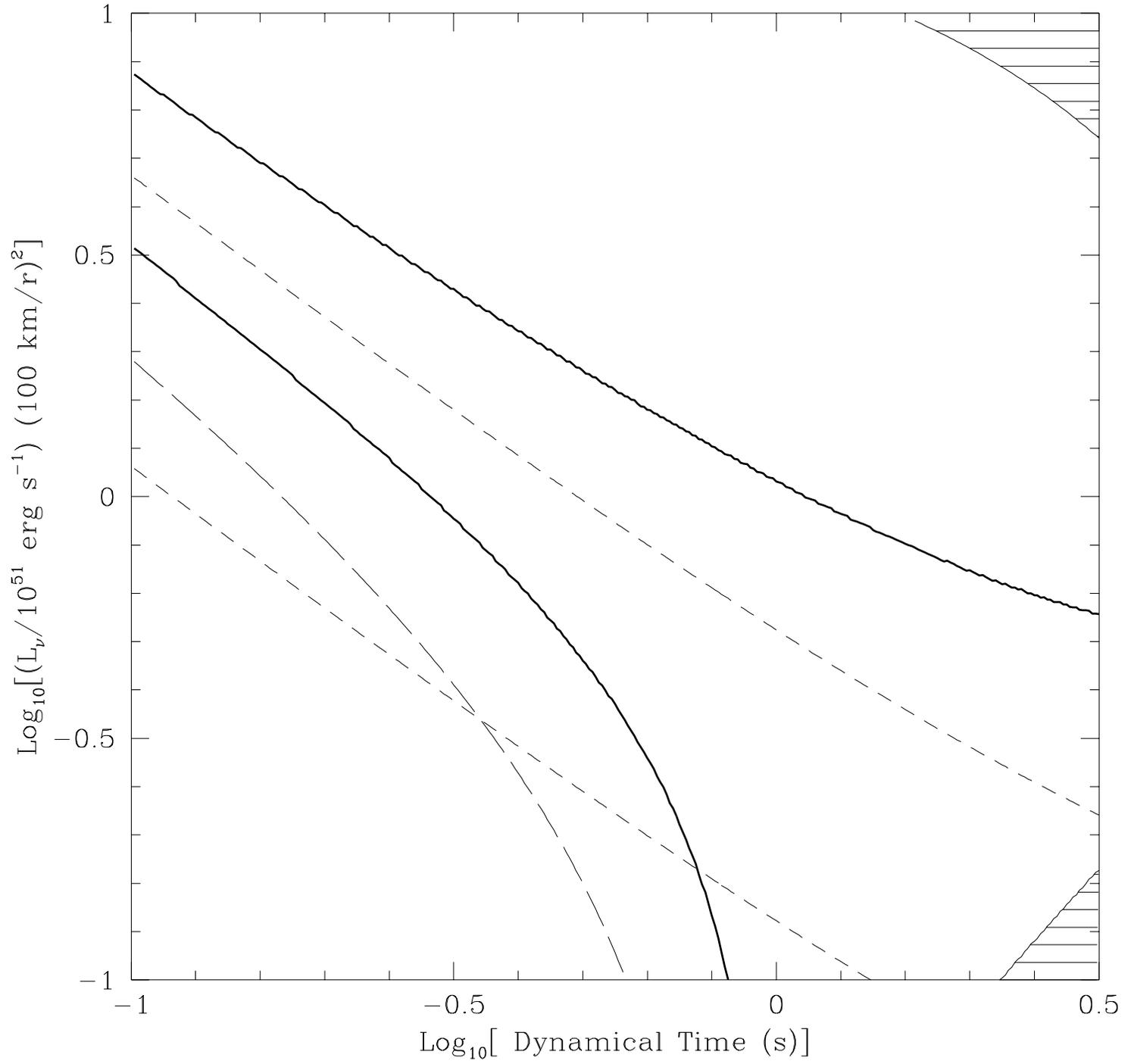
\end{document}